# Promoting Multimedia in Physics Teaching Through the Flipped Classroom in Pre-Service Education

*Vera Montalbano*
*University of Siena, Italy*

*Abstract*
A flipped classroom was introduced in pre-service education in order to promote MM and the quality of their use in the teaching and learning process in laboratory. The course, Physics Lab Didactics, promoted active learning through the direct experience of young teachers. The flipped classroom approach was presented and discussed with the aim of clarifying the teaching process and in-training teachers were invited to explore how to implement this methodology in a class after their experience. Different open source or free software were proposed (*Audacity*, *Algodoo*, *GeoGebra* and *Tracker*) with some examples of their use in physics teaching. The discussion on their implementation as tools in physics laboratory was postponed to the last lesson of the course. This pilot study shows a great potentiality of the flipped classroom in pre-service education and it indicates that the use of MM in laboratory learning process can be improved following this approach.

## 1 Instruction

The increasing role of technology in contemporary society, as well as the rapid advancement of technology types and uses, requires major changes to methods of teaching. The use of a flipped classroom approach in a pre-service course can promote active learning in young teachers, enhance the critical thinking and obtain the maximum use of student-faculty time together (Keengwe, 2014). This constructivist approach to teaching is an effective means of student-centered collaboration. The benefits are many, such as to maximize the learners' capacity to engage in small group discussion, project based learning, or problem solving tasks (Larcara, 2014). Moreover, the face-to-face classroom time can be used for peer collaboration, inquiry, and project-based learning (Dickenson, 2014).
Multimedia tools are rarely proposed in pre-service formation of physics teachers in Italy.
A pilot study on implementing the use of multimedia is the latest in a series of actions for transforming teacher practice through professional development (Montalbano, Benedetti, Mariotti, Mariotti, and Porri, 2012). The next section describes the context in which the flipped classroom was introduced. In the following one, its implementation and the materials are reported. The results are reported and discussed in the last section.

## 2 Concept of the Teaching Experience

In the pre-service education of Italian physics teachers, the introduction of multimedia tools is usually absent or marginal. Recently, a pilot study with the aim to change this situation was realized by mean of a flipped classroom. The implementation allowed to foster the MM presented to young teachers and to produce educational materials ready for the use at school. Another relevant feature of this teaching experience was the possibility of a direct involvement for teachers in a flipped classroom, in order to clarify how this new methodology works. Let us start by describing the context in which the pilot study was realized.

### 2.1 The Pre-Service Education

In the last fifteen years, pre-service education drastically changed in Italy. The main steps in this transformation are shown in table 1. From the complete absence of a pre-service formation, we arrived to an annual advanced course named Formative Active Training (*Tirocinio Formativo Attivo*, i.e. TFA) and the process is not still ended. It seems that the next TFA could be the last one before a new reform in teachers education.

The first introduced pre-service education was a biennial advanced school where only basic ICT tools were proposed in disciplinary context. In the TFA course, the focus on direct training experience led to an increase in the time devoted to training at school and a consequent reduction in the available time for developing disciplinary and educational competencies. Since the use of MM is not widespread in the Italian school, it in very difficult to change this situation and to foster their integration in current didactics.

|  | pre-service edu Adv. course | Adm exam | Pre-service training | Teaching Qualification |
|---|---|---|---|---|
| **Before 1999** | none | no | none | Professorships competitive examination |
| **1999 - 2009** | SSIS biannual | yes | 290 hours/year | Exam for teaching qualification (written and oral exam) |
| **2012 - 2016** | TFA annual | yes | 475 hours | Exam for teaching qualification (final report on training and oral exam) |

*Tab. 1. Teacher education in Italy in last decades.*

In this context, the opportunity of increasing the knowledge in MM through an active involvement in a flipped classroom emerged. Moreover, it was even the only possibility of dealing with this issue in the little time available.

## 2.2 Participants

The flipped classroom was implemented in a course on Physics Lab Didactics where a special care was given in introducing active learning (Bonwell and Eison, 1991), proposing examples focused on behavior that can facilitate or inhibit it in lessons and especially in lab designing and execution. Participants had usually few or no experience in phys lab and active learning was implemented in disciplinary contents (Montalbano and Benedetti, 2013). Focus was on teachers engagement, how to work in group, cooperative learning (Cuseo, 1992) and so on. In laboratory everything can go in a wrong or unexpected way, giving the opportunity to highlight how to interact with students (a lot of good examples usually happened in lab and how to manage them was discussed together).

Due to the admission test to the TFA and to the territorial distribution of Tuscan universities, only a limited number of students were enrolled for teaching Mathematics and Physics in secondary school of $2^{nd}$ grade.

Participant enrolled in the last TFA were four, two with a Physics degree and the others with a degree in Mathematics. Moreover, three of them were qualified for teaching Mathematics and Science in secondary school of $1^{st}$ grade with the previous TFA.

## 3 The Flipped Classroom

The main purpose of the action was to foster the use of MM in physics education through a direct involvement of young teachers in specific tasks. The idea was to show them that to learn by themselves how a MM tool works, design and develop new effective materials for the use in class was possible even in few time. Despite the choice of the flipped classroom approach was imposed by time constraints, the reason for introducing this methodology, the main attentions, the pros and cons were discussed in an initial participate lesson.

## 3.1 MM Proposals

The first attempt of introducing a MM tool for physics education in pre-service formation was done in the previous TFA course (about ten participants) by introducing some example of using Audacity by a participate lesson. In this case, the assessment was disappointing: very weak engagement for teachers and no production of educational materials.

For the flipped classroom were proposed four open source software (*Audacity*, *GeoGebra*, *Algodoo*, *Tracker*). Few words were spent for introducing them (where to download, why each software can be useful, and so on) and presenting the proposed materials (1-2 examples in PER for each tool).

In particular, for *Tracker* an exploration of kinematics and dynamics of a game (Rodrigues and Simeão Carvalho, 2013) was proposed. For *Audacity* and *Algodoo*, the examples were taken from the 18th Edition of the Multimedia in Physics Teaching and Learning Conference. An interdisciplinary learning path on sound and noise (Montalbano, 2014) for the former, a simulation for the process of diffusion (da Silva, Junior, da Silva, Viana, and Leal, 2014) for the latter.

Finally, for *GeoGebra* no additional material was proposed because young teachers attended to another course in TFA where this software was widely utilized for generating conjectures in dynamic geometry (Baccaglini-Frank and Mariotti, 2010; Leung, Baccaglini-Frank, and Mariotti, 2013) and used in extensive way in TFA in a Math didactics course.

### 3.2 Tasks

The goal of maximizing the exploration with the proposed MM tools was achieved by assigning the following tasks to each student:
- Choose a different MM tool (promote cooperation in organizing the task);
- Learn to use the MM tool (require active learning);
- Design a learning path related to a physics lab topic where the MM tool can display an effective help in the learning process and implement it in details (promote active learning in teaching skills);
- Share his/her learning path by a peer communication with explicit remarks on methodological choices, strengths and weaknesses in the use of the software (maximum dissemination of MM and learning paths).

The peer sharing was realized in a participate lesson through an interactive presentation and written worksheets for each proposed learning path were mandatory.

### 4 Results and Remarks

The main goal achieved with the flipped classroom was to interact with teachers who had already actively used the software in an educational context. The materials produced were discussed both from the point of view of the learning process that to highlight the characteristics of the software. All materials were shared in a cloud storage service whose link was created and shared by teachers.

Each teacher presented a different learning path: a study of a falling body using the software as tool for measuring and fitting from a video (Letizia), a simulation of different situation in gravity field for verifying a Galileo's theorem (Riccardo), an analysis on different sounds for understanding their features and introducing to resonance (Stefano) and an interdisciplinary path on the Huygens' pendulum (Mauro).

### 4.1 On Studying Dynamics with *Tracker*

The proposal is a standard laboratory experiment for exploring the dynamics of a falling body but it was performed by analysing a video by using *Tracker*.

In fig. 1 a screen from the analysis is shown. The video showed a little ball falling and rebounding several times like it is shown in the complete analysis of given in fig. 2, where the vertical position, velocity and acceleration measured from the frames are shown in function of time.

Moreover, first falling data were selected and used for fitting the acceleration of gravity *g* in two different ways: by using a parabolic function for the position and a linear regression for the velocity. Both results are shown in fig. 2.

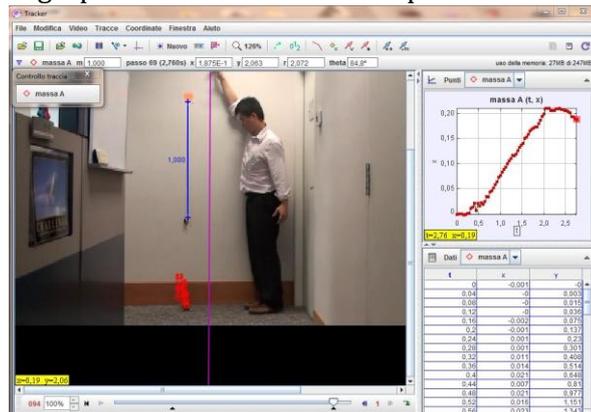

*Fig. 1. The proposal was based on a video found online of the fall and subsequent rebounds of a little ball.*

The discussion focused on the use of the MM tool, about some difficult in measuring, the evaluation of uncertainties and the quality of the fitting tool.

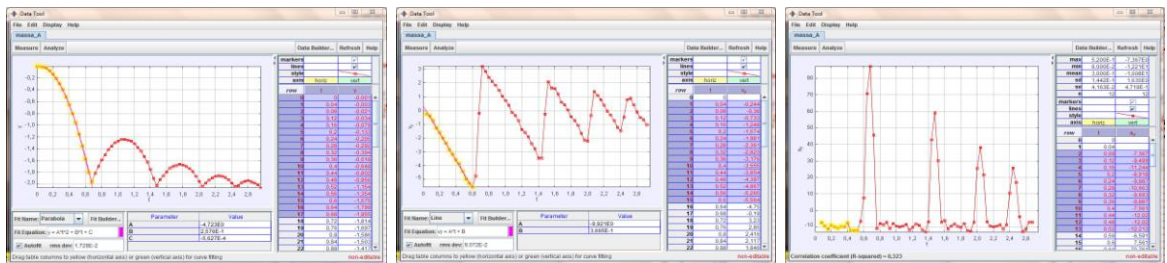

*Fig. 2. The figure shows program screens for some activities proposed in the learning: the vertical position in the time (on the left), the vertical velocity (on the center) and the vertical acceleration (on the right). In each screen the data for the initial falling are selected (in yellow) and shown on the right. Moreover, the fit for the selected data is shown below for the position and velocity.*

Finally, the possibility of studying collisions, especially from point of view of energy, was explored.

### 4.2   Verify a Galileo's Theorem with *Algodoo*

"Algodoo is a program of physical simulation, very similar to a game. Objects of any shape are easily created, with the difference they suddenly become "real" in the physical world simulation, for example, as soon as created, we will see them fall to the ground" (description of the software by Riccardo to others).

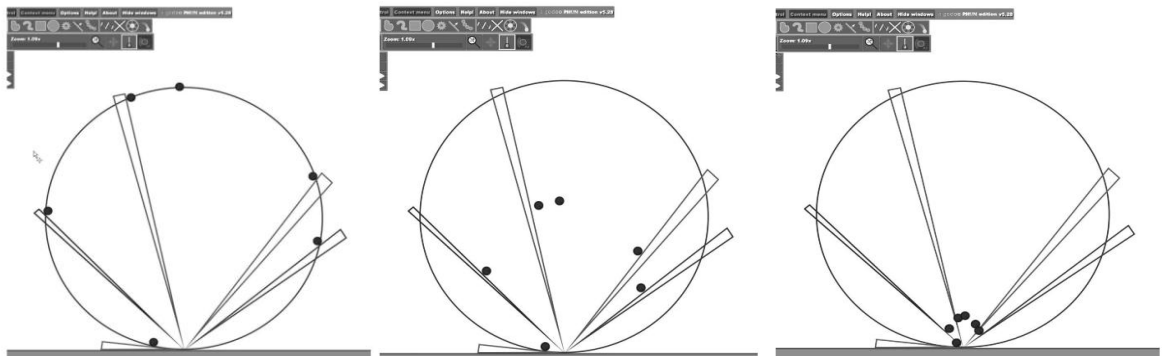

*Fig. 3. The figure shows program screens for the activity proposed for "discovering" Galileo's Theorem.*

The proposal was a simulated experiment for introducing the Galileo's theorem on the descent time along two-chord paths in a circle.

The use of the software in physics and mathematics education was promoted by Riccardo in this way: "How many interesting experiments are never realized because the result not worth the time spent to prepare them! The simulation allows us to recover with ease, precision and relative speed at least some of these experiences. Thus, I propose this activity: easy to make, it shows spectacular effect that leaves amazed those who do not already know the result, useful for concepts that recalls and questions which can be posed, and it bypasses exercises that can complete it, whose realization would be very hard-working, . . . and it was based only on falling bodies!"

A limit of the software was outlined in the lack of the possibility of introducing a variable field of force. This aspects prevents its use for simulation of celestial mechanics or motion of electric charges. The teacher tested another software with this possibility (*Step*) but he encountered too many bags and his opinion was that it is not still ready for a direct use at school, on the contrary *Algodoo* is fully ready.

### 4.3 A Learning Path on Sound and Music with *Audacity*

The learning path was centred on the characterization of sounds through a software audio editor. Students receive an audio composed of various tracks, synthesized by the teacher and characterized by different frequencies and different waveforms (sine, square, saw tooth, see fig. 4). Students must characterize quantitatively track by track and qualitatively as a set of tracks: study of waveform (time domain) of each track (measuring amplitude, sampling frequency, frequency), study of the spectrum of each track (frequency domain) finding the frequency of the peak and the peak width. The next step is an overall analysis from feelings derived from the combination of different tracks: which of these tracks sound harmoniously, which are dissonant and which look very similar? Is there any law that binds the frequencies of the sounds between them similar? And among those harmonics?

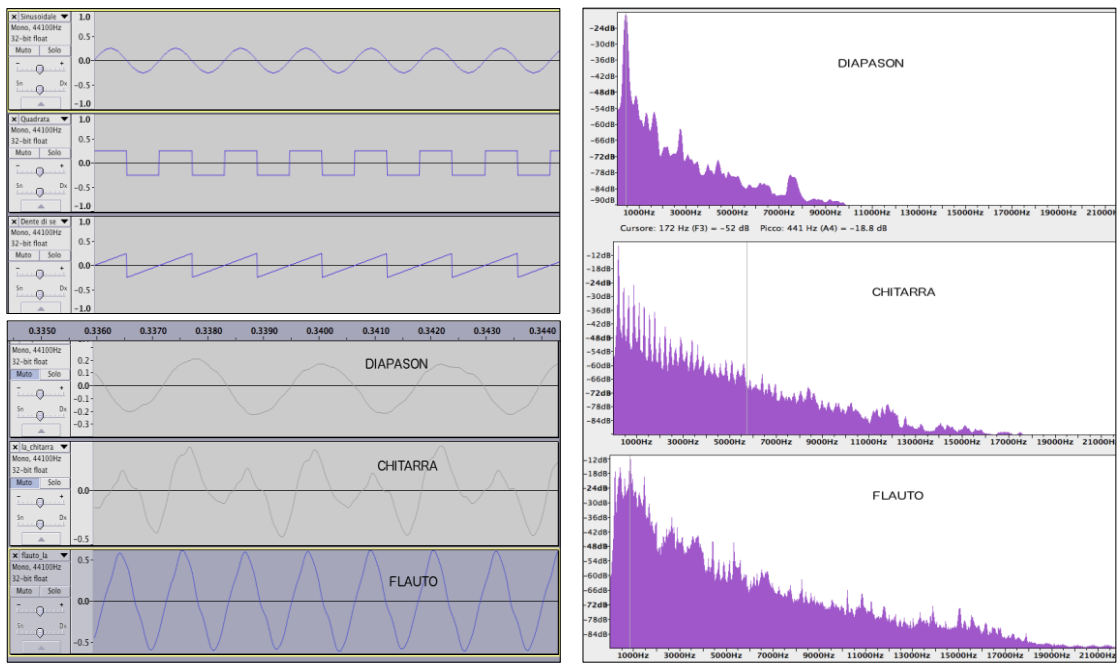

*Fig. 4. The figure shows program screens for the initial activities. The sound tracks prepared by the teacher (on the top of left side) and obtained by recording sounds from a diapason, a guitar and a flute (on the bottom of left side), whose spectre are shown on the right side.*

The following activity is the study of recordings of individual notes produced by various instruments (guitar, flute) and mechanical oscillators (diapason), whose sound tracks and spectra are shown im fig. 4. What tools emit sounds more "simple"? And which ones more compounds?

The last activity is the study of human voice by recording single vocal and singing a note and compare the spectrum with the case in which sound enters in an open tube (Hula Hoop). All spectra are shown in fig. 5.

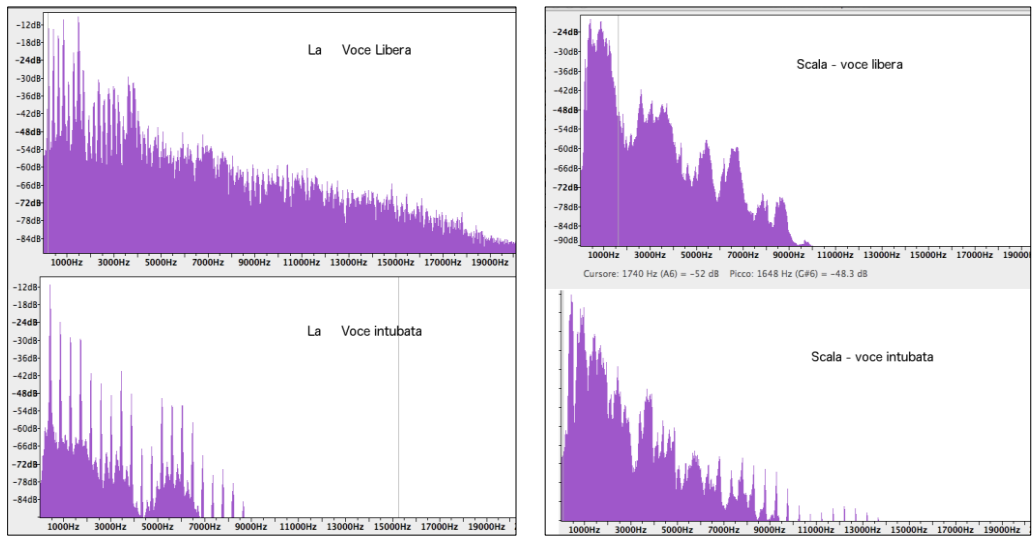

*Fig. 5. Spectrum obtained by analysing a free human voice pronouncing a vocal (on the top of left side) and inside an open tube (on the bottom of the left side). Spectrum of a human voice singing a scale (top of the right side) and the same sound inside an open tube (on the bottom).*

The spectra showed in fig. 5 were realized by the teacher during the peer presentation with with the aid of his daughter (about ten year old) demonstrating the full involvement in the research of useful and easy activity really realisable at school.

### 4.4 An Interdisciplinary Learning Path on the Heygens' Pendulum

The last proposal was the more articulated, interdisciplinary and interesting. The purpose was to compare a pendulum with a Heygens' pendulum from theoretical and experimental point of view. Thus, it was necessary to construct and verify the isochronism of a Heygens' pendulum but also to demonstrate all interesting property of a cycloid function in a high school class.

In order to verify all property of a cycloid, dynamical demonstrations were constructed by using *GeoGebra* (see a frame of the demonstration of the involute in fig. 6).

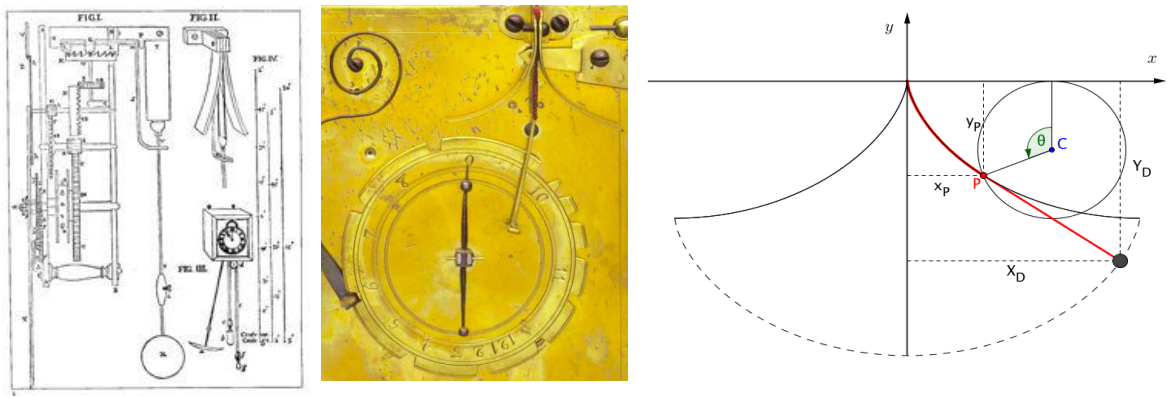

*Fig. 6. The original figure (Heygens 1673) with the proposed pendulum mechanism by Heygens (on the left), details of Huygens' Chops in an ancient clock (on the center), cycloid function and its involute from a frame of the dynamics demonstration constructed by GeoGebra (on the right).*

The Heygens' pendulum is an historic example of relation between mathematics, physics and technical progress which introduces to complex relations that bounds science and society. In figure 6, a picture of the pendulum from the original proposal of Huygens and a details of Heygens' chops following a cycloid shape in an ancient clock are shown.

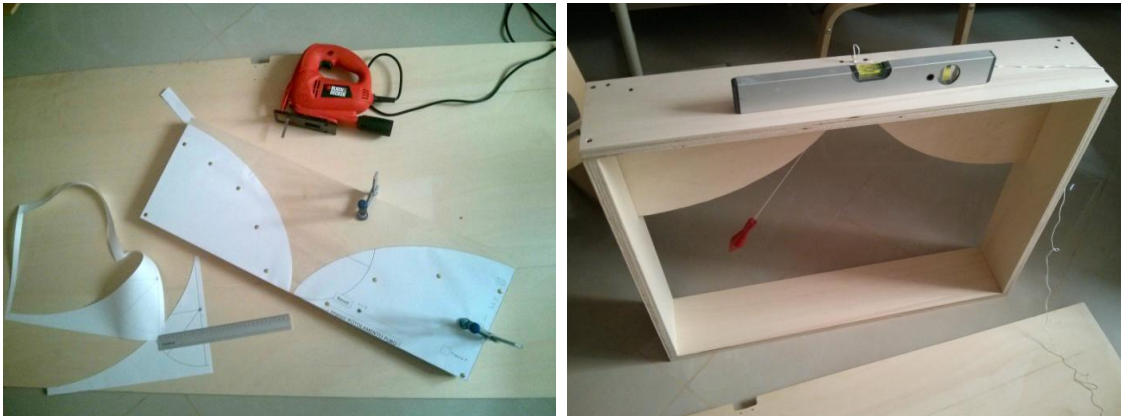

*Fig. 7. Two steps in the realisation of the Heygens' pendulum: the use of printed cycloid from Geogebra (on the left side) and the levelling of the device (on the right side).*

*GeoGebra* allows to generate a print of a cycloid adapted to shape Heygens' chops in wood (see fig. 7 for relevant steps in the construction of the pendulum).
The final device was optimized for video analysis with *Tracker* choosing an effective shape and colour for the pendent mass. An example of measurement is given in figure 8.

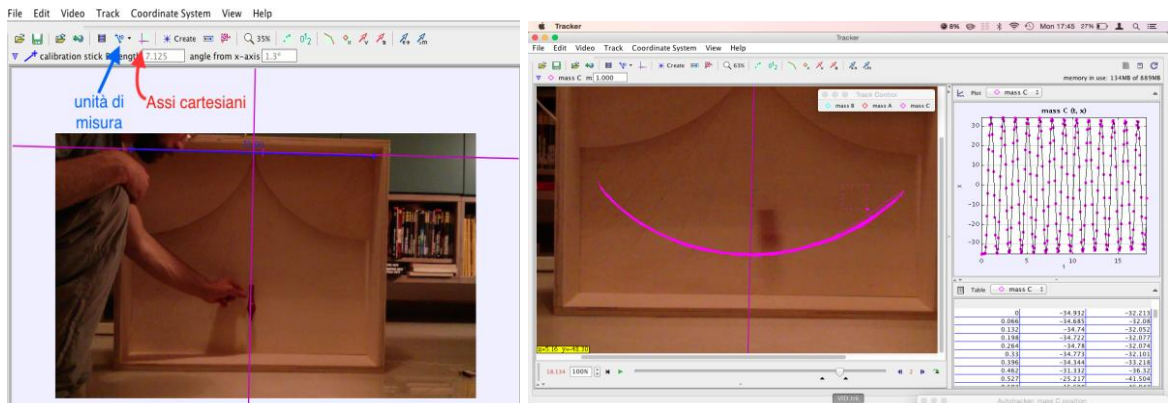

*Fig. 8. The measure with Tracker: Positioning the axis in the initial frame (on the left) and a sampling of measures (on the right).*

The devise was very versatile, allowing to perform measurement with pendulum and Heygens' pendulum. It was possible to move the two pendula simultaneously and observe almost no difference with small initial angles or a defined asynchrony when initial angles are larger.
The most interesting aspect from the experimental point of view was the sensibility of the device when used with *Tracker*. The isochronism of a Heygens' pendulum was tested for initial angles in the range 15°- 50° and the measured period resulted to be always the same: 1.35±0.02 s.

### 4.5   Remarks and Conclusions
The main goal achieved with the flipped classroom was the excellent feedback from teachers. Direct assessment was performed during lesson time, laboratory and MM products were shared together with new ideas and educational materials, maximizing the available time.

Discussions on the use of MM during arose during physics laboratory. A creative use of GeoGebra in the data analysis for measuring the time of discharge of an RC circuit emerged about the visualization of the best fit process (Montalbano and Sirigu 2015).

Teachers developed good ideas which led to realizing well designed specific learning paths. The success in the teaching experience depends from a careful choice of the initial materials proposed for the flipped classroom and a well-established student-teacher relationship (trust gained in the previous TFA course) in order to avoid the first and not so hidden thought that students usually have as soon as a flipped classroom is proposed: the teacher is lazy and escapes its duty and role.

Another relevant aspect is the little number of participants and the fact that they were graduate students really interested in a professional development of their skills. Moreover, the chosen topic, MM tools in Physics education, was very suitable for the raised interest and the richness of materials that can be found. Last but not least, participants were excellent students, forming a very good working group, and hopefully will become excellent teachers.